\documentclass{midl} 

\usepackage{mwe} 
\jmlrproceedings{MIDL}{Medical Imaging with Deep Learning}
\jmlrpages{}
\jmlryear{2019}
\jmlrworkshop{MIDL 2019 -- Extended Abstract Track}

\title[Rib Labeling and Segmentation using Mask R-CNN]{Sequential Rib Labeling and Segmentation in Chest X-Ray using Mask R-CNN}

\midlauthor{\Name{J\"oran Wessel}\nametag{$^{1,2}$}\\
\addr $^{1}$ Philips Research, Hamburg, Germany \\
\addr $^{2}$ Institute of Medical Informatics, Universit\"at zu L\"ubeck, L\"ubeck, Germany\\
\Name{Mattias P. Heinrich\nametag{$^{2}$}}\\
\Name{Jens {von Berg}\nametag{$^{1}$}}\\
\Name{Astrid Franz\nametag{$^{1}$}}\\
\Name{Axel Saalbach\nametag{$^{1}$}}
}

\begin{document}

\maketitle

\begin{abstract}
Mask R-CNN is a state-of-the-art network architecture for the detection and segmentation of object instances in the computer vision domain. In this contribution, it is used to localize, label and segment individual ribs in anterior-posterior chest X-ray images. 

For this purpose, several extensions have been made to the original architecture, in order to address the specific challenges of this application. This includes the use of rib specific networks, facilitating dedicated anchor boxes sampled from a training set, as well as a sequential processing of all ribs. Here, the segmentation result of the upper neighbor rib is used as additional input to the network.

This approach is the first addressing both rib segmentation and anatomical labeling in chest radiographs. The results are comparable or even better than existing methods aiming only at segmentation.
\end{abstract}


\section{Introduction}

The identification of ribs has many applications in chest radiography. Ribs may obscure important findings in the lung parenchyma, why rib shadows can be either excluded from the automatic analysis \cite{Candemir2016} or suppressed from the image \cite{Berg2016} to minimize their impact. In contrast to rib segmentation, rib labeling is required in chest X-ray quality assessment. Here, counting the ribs visible in the lung field, is a standard procedure to assure proper inhalation state~\cite{Mader2018}. But also other applications like automatically localizing findings from a report or establishing correspondence between follow up images may benefit from a rib segmentation and labeling method.
No method published so far does achieve both rib segmentation and rib labeling.\\

Mask R-CNN \cite{He2017} is a convolutional neural network for simultaneous object detection and segmentation (i.e. instance segmentation), developed for real time video processing. In the following, we discuss an extension of the Mask R-CNN algorithm and its application to chest X-ray analysis.

\section{Data and Method}
In this work we use the dataset described in~\cite{Berg2016} which contains 174 posterior-anterior chest X-ray images. The ribs in this dataset were contoured by hand earlier in order to evaluate a bone suppression method, so that only the visible shadows of the ribs are properly annotated. For each X-ray image, the analysis was restricted to the ribs 1 - 9 (which were consistently visible). For the analysis, the images and annotation masks are down sampled to a pixel spacing of approximately 1~mm. This corresponds to an image size of less than $ 500\times500 $ pixel. Additionally, the training image set was augmented by using affine transformations.

For our experiments we extend a publicly available implementation\footnote{https://github.com/multimodallearning/pytorch-mask-rcnn}, with a ResNet50~\cite{He2016}
and a Feature Pyramid Network~\cite{Lin2017} as a feature extractor.
The first extension of our method includes the use of dedicated anchor boxes. Instead of independently recognizing and evaluating Regions of Interest with the Region Proposal Network, the network determines shifts on the basis of anchor boxes which correspond to typical rib locations (without Non-Maximum Suppression). 
In this experiment, the anchor boxes have been estimated using the Mean Shift algorithm~\cite{Comaniciu2002}. Therefore, 30 cluster boxes were computed from all ground truth bounding boxes of the entire dataset for all labels. In order to compensate for different sizes of the X-ray images, the ground truth boxes are normalized with respect to the size of the respective image. 

Due to the high self-similarity of adjacent ribs, reliable rib labeling is a challenging task. Therefore, as a second extension, a sequential processing scheme is introduced, similar to the method from \citet{lessmann2019iterative}. Here, separate networks are trained for each rib so that the networks always segment and classify the same rib for the shifted cluster boxes. Using a ResNet architecture for X-ray image analysis, the image data is typically replicated across the three input channels (RGB). In this setup however, for ribs 2 to 9, the output of the segmentation of the above rib is used for the third channel. The first two channels are filled with the X-ray image in gray values. The idea of this extension is that the additional information of the upper neighbor rib makes the detection of the current rib more precise and reduces false detections.

\section{Results}
For an evaluation of the proposed architecture, a five-fold cross-validation was performed. In \figureref{fig:segmentation} the outcome of the algorithm is depicted for one exemplary case. The high agreement between the ground truth annotations (left panel) and the obtained results (right panel) is also reflected in a quantitative evaluation for the predicted bounding boxes and the segmentations.

\begin{figure}[h]
  \floatconts
  {fig:segmentation}
  {\caption{Application of the adapted Mask R-CNN algorithm to a chest X-ray image. The ground truth is shown on the left and the prediction on the right. For illustration purposes, next to the segmentation, also the bounding boxes and the rib labels are given.}}
  {\includegraphics[width=0.95\linewidth]{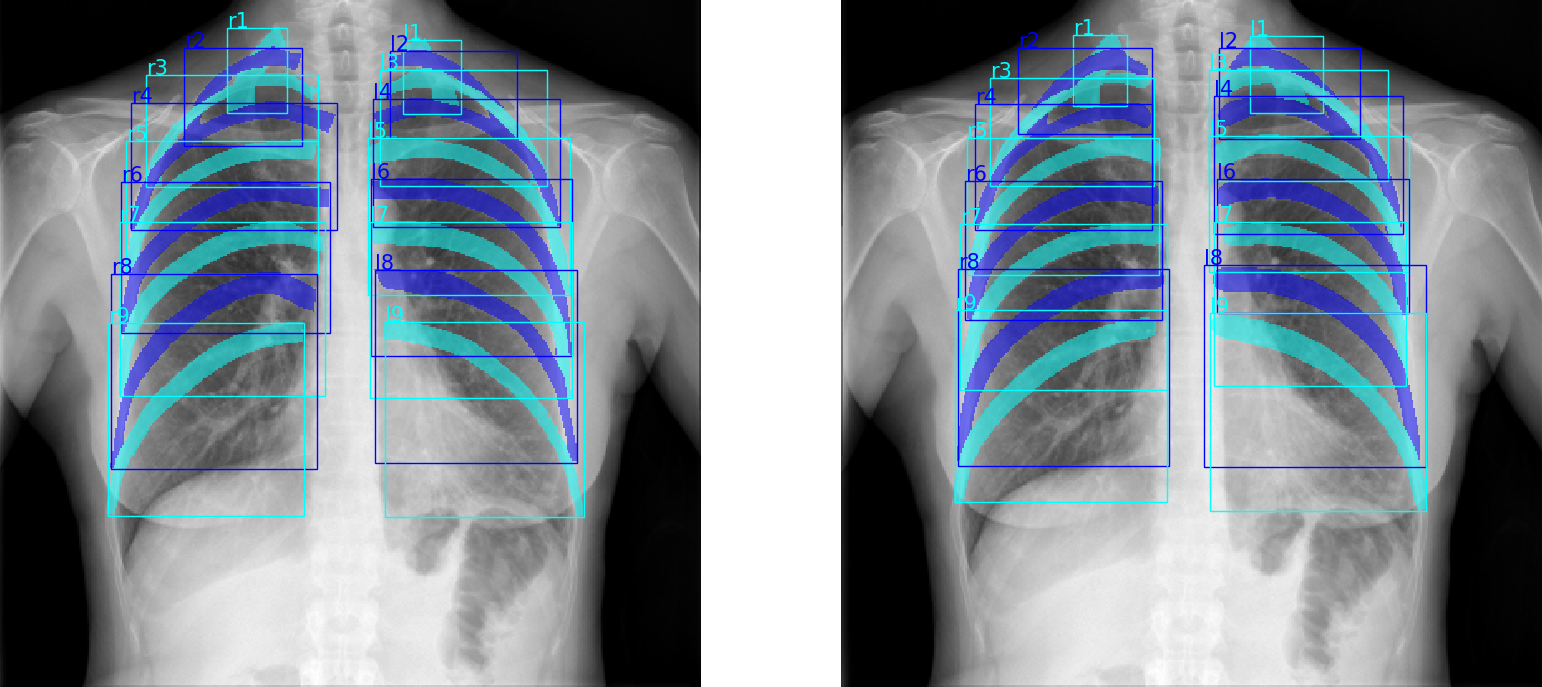}}
\end{figure}

The results for the detection of the bounding boxes, as well as for the instance segmentation of the ribs are given in Table~\ref{tab:results}.
With the proposed architecture, mean Dice values of 0.846 for the bounding boxes and 0.733 for the segmentation are achieved
Compared to the original implementation of the algorithm, this is an improvement of more than $5\%$ and $23\%$, respectively.

\begin{table}[h]
	\floatconts
	{tab:results}
	{\caption{Dice coefficients for rib detection and rib instance segmentation in a five-fold cross-validation. The dataset covers 174 posterior-anterior chest X-ray images~\cite{Berg2016}}}%
	{\begin{tabular}{lcc}
		& \bfseries Detection & \bfseries Segmentation\\
			Left & $ 0.841 \pm 0.126 $ & $ 0.732 \pm 0.207 $\\
			Right & $ 0.850 \pm 0.104 $ & $ 0.734 \pm 0.211 $
	\end{tabular}}
\end{table}

Previously, the problem of rib detection has been addressed by Candemir et al. \cite{Candemir2016}, using atlas-based methods, using accuracy, sensitivity and specificity as evaluation criteria. With values of $0.95$, $0.82$ and $0.98$ compared to $0.86$, $0.75$ and $0.92$, our algorithm achieves better results in all categories.

\section{Conclusion}
In this contribution we presented the first approach for simultaneous rib detection and segmentation in thoracic radiography.
The introduced enhancements result in a considerable improved detection rate and segmentation accuracy compared to the original Mask R-CNN approach. Furthermore, it outperforms existing state-of-the-art techniques using atlas-based registration, while providing very fast run-times that enable realtime analysis.

\newpage

\bibliography{wessel19}

\begin{thebibliography}{8}
\providecommand{\natexlab}[1]{#1}
\providecommand{\url}[1]{\texttt{#1}}
\expandafter\ifx\csname urlstyle\endcsname\relax
  \providecommand{\doi}[1]{doi: #1}\else
  \providecommand{\doi}{doi: \begingroup \urlstyle{rm}\Url}\fi

\bibitem[Candemir et~al.(2016)Candemir, Jaeger, Antani, Bagci, Folio, Xu, and
  Thoma]{Candemir2016}
Sema Candemir, Stefan Jaeger, Sameer Antani, Ulas Bagci, Les~R Folio, Ziyue Xu,
  and George Thoma.
\newblock Atlas-based rib-bone detection in chest {X}-rays.
\newblock \emph{Computerized Medical Imaging and Graphics}, 51:\penalty0
  32--39, 2016.

\bibitem[Comaniciu and Meer(2002)]{Comaniciu2002}
Dorin Comaniciu and Peter Meer.
\newblock Mean shift: A robust approach toward feature space analysis.
\newblock \emph{IEEE Transactions on Pattern Analysis and Machine
  Intelligence}, 24\penalty0 (5):\penalty0 603--619, 2002.

\bibitem[He et~al.(2016)He, Zhang, Ren, and Sun]{He2016}
Kaiming He, Xiangyu Zhang, Shaoqing Ren, and Jian Sun.
\newblock Deep residual learning for image recognition.
\newblock In \emph{Proceedings of the IEEE Conference on Computer Vision and
  Pattern Recognition}, pages 770--778, 2016.

\bibitem[He et~al.(2017)He, Gkioxari, Doll{\'a}r, and Girshick]{He2017}
Kaiming He, Georgia Gkioxari, Piotr Doll{\'a}r, and Ross Girshick.
\newblock Mask {R-CNN}.
\newblock In \emph{Computer Vision (ICCV), 2017 IEEE International Conference
  on}, pages 2980--2988. IEEE, 2017.

\bibitem[Lessmann et~al.(2019)Lessmann, van Ginneken, de~Jong, and
  I{\v{s}}gum]{lessmann2019iterative}
Nikolas Lessmann, Bram van Ginneken, Pim~A de~Jong, and Ivana I{\v{s}}gum.
\newblock Iterative fully convolutional neural networks for automatic vertebra
  segmentation and identification.
\newblock \emph{Medical Image Analysis}, 53:\penalty0 142--155, 2019.

\bibitem[Lin et~al.(2017)Lin, Doll{\'a}r, Girshick, He, Hariharan, and
  Belongie]{Lin2017}
Tsung-Yi Lin, Piotr Doll{\'a}r, Ross~B Girshick, Kaiming He, Bharath Hariharan,
  and Serge~J Belongie.
\newblock Feature pyramid networks for object detection.
\newblock In \emph{CVPR}, volume~1, page~4, 2017.

\bibitem[Mader et~al.(2018)Mader, von Berg, Fabritz, Lorenz, and
  Meyer]{Mader2018}
Alexander~Oliver Mader, Jens von Berg, Alexander Fabritz, Cristian Lorenz, and
  Carsten Meyer.
\newblock Localization and labeling of posterior ribs in chest radiographs
  using a {CRF}-regularized {FCN} with local refinement.
\newblock In \emph{International Conference on Medical Image Computing and
  Computer-Assisted Intervention}, pages 562--570. Springer, 2018.

\bibitem[von Berg et~al.(2016)von Berg, Young, Carolus, Wolz, Saalbach,
  Hidalgo, Gim{\'e}nez, and Franquet]{Berg2016}
Jens von Berg, Stewart Young, Heike Carolus, Robin Wolz, Axel Saalbach, Alberto
  Hidalgo, Ana Gim{\'e}nez, and Tom{\'a}s Franquet.
\newblock A novel bone suppression method that improves lung nodule detection.
\newblock \emph{International Journal of Computer Assisted Radiology and
  Surgery}, 11\penalty0 (4):\penalty0 641--655, 2016.

\end{thebibliography}
\end{document}